# Structural and Magnetic properties of $Ge_{0.5}Mn_{0.5}Co_2O_4$ using neutron diffraction

Pooja Jain[1], Benny Schundelmier[2], Chin-Wei Wang[3], Poonam Yadav[4], Kaya Wei[2], N. P. Lalla[1], Shivani Sharma[2,5*]

[1]*UGC-DAE Consortium for Scientific Research, Indore, Khandwa Road, MP India.*
[2]*National High Magnetic Field Laboratory, Tallahassee, FL 32310, United States.*
[3]*National Synchrotron Radiation Research Center, Hsinchu 300092, Taiwan.*
[4]*Center for Integrated Nanostructure Physics, Institute for Basic Science, Suwon 16419, Republic of Korea.*
[5]*Kazuo Inamori School of Engineering, Alfred University, Alfred, New York 14802, United States.*

**Abstract**

The structural and magnetic properties of $Ge_{0.5}Mn_{0.5}Co_2O_4$ (GMCO) have been investigated in detail utilizing neutron powder diffraction (NPD), x-ray diffraction (XRD), DC magnetometry, and heat capacity analysis and compared with $GeCo_2O_4$. Despite both compounds exhibiting a cubic structure at room temperature, a substantial difference on low temperature structural properties have been observed for $Ge_{0.5}Mn_{0.5}Co_2O_4$, indicating the effect of Mn substitution on crystal structure. The magnetic and heat capacity data reveal a ferrimagnetic ordering around 108 K in GMCO. A minor secondary phase is presented in GMCO which undergoes long range AFM ordering at further lower temperatures. This secondary phase remains undetected in XRD, neutron powder diffraction reveals its presence, exhibiting nearly identical lattice parameters. Furthermore, the analysis of heat capacity data indicates a broadening of the high-temperature transition, attributing to the short-range correlation persisting up to higher temperatures. The estimated magnetic entropy amounts to only 63% of the value expected with Mn and Co ions. The missing entropy is likely linked with the short-range magnetic correlations persisting well above the transition temperature. Using Rietveld analysis of neutron powder diffraction data at room temperature, cation distribution at the A and B sites has been estimated in GMCO. Magnetic structures are also confirmed in the main phase as well as in the secondary phase using NPD analysis. The high-temperature transition corresponds to the ferrimagnetic ordering of A and B site cations in the main phase. A collinear ferrimagnetic arrangement of A and B site spins aligned parallel to *c* axis is observed. The average values of A and B site moments in the ferrimagnetic phase at 5 K are 2.31(3) and 1.82(3) $\mu_B$, respectively, with the temperature dependence of moments following the expected power law behaviour. The low-temperature ordering is attributed to the antiferromagnetic ordering of B site ions associated with the secondary phase, something similar to $GeCo_2O_4$.

**Introduction**

Over the past several decades, the exploration of frustration in magnetic systems has been a subject of intense interest, primarily driven by the captivating magnetic states, such as spin glass or spin liquid in strongly correlated electron systems [1–10]. One of the most studied origins of frustration in magnetic systems is the geometrical arrangement of first nearest neighbour antiferromagnetic interactions in triangular (2D), tetrahedral (3D), pyrochlore, or Kagome lattices [5,10–14]. This discussion particularly focuses on geometrically frustrated systems, exemplified by Co-based cubic spinel structures ($ACo_2O_4$), where the Co atoms are arranged in alternate planes of triangular and Kagome lattice [3,4,13,15–20]. Due to corner-sharing tetrahedra of B-site cations forming pyrochlore lattice, spinel exhibit geometrical frustration. Frustration in spinel compounds can be attributed to the Jahn-Teller effect and spin-orbit interactions, leading to phenomena like orbital glass and liquid states [8,18,21–23]. Moreover, spinel oxides, distinguished by different cation distributions, are broadly classified into two categories: Normal Spinel and Inverse Spinel. In normal spinel, A cations occupy tetrahedral sites, and B cations occupy octahedral sites, following the cubic space group $Fd$-$3m$. Each formula unit features eight tetrahedral and four octahedral sites. A notable example of a normal spinel is $GeCo_2O_4$ (GCO) [19]. However, in the inverse spinel configuration, all A cations and half of the B cations occupy octahedral sites, while the remaining B cations occupy tetrahedral sites, such as the case in $MnCo_2O_4$ (MCO) [24,25]. Significant research has been conducted on both GCO and MCO, shedding light on their structural characteristics and magnetic properties. GCO has garnered significant attention in research due to its unique electronic and magnetic ground state featuring octahedral $Co^{2+}$. This state is characterized by a high-spin $3d^7$ configuration with S=3/2, L=3, yet is more accurately described as a Kramer's doublet with $J_{eff}$=1/2. The presence of orbitally degenerate $t_{2g}^5$ states in the high-spin octahedral $Co^{2+}$ leads to substantial spin-orbit coupling, resulting in a pronounced single-ion anisotropy, a distinctive trait for a $3d$ transition metal. Below its Néel temperature ($T_N \cong 21$ K), GCO exhibits antiferromagnetic ordering with a characteristic wave vector k = (1/2,1/2,1/2) [3,4,11–13,15,19,21,26]. This ordering is accompanied by a structural phase transition from cubic to tetragonal symmetry [19]. Notably, neutron studies conducted by multiple research groups offer a cohesive understanding of the spin ordering in GCO [3,12,21,26]. On the other hand, MCO has been studied for its remarkable magnetic properties and colossal magnetoresistance behaviour [24,25]. The compound MCO is notably intriguing, particularly below 130 K, where magnetic hysteresis curves exhibit unconventional behaviour, which was suggested to be linked with the irreversible domain wall movements that overcome the anisotropy field below 130 K [24]. However, the unusual magnetic hysteresis as observed in the case of MCO is debatable and needs to be confirmed to exclude the artifacts including the trap field effect on the magnetic response on a single crystalline sample. Moreover, the effect of pressure on the relative magnetic susceptibility of MCO is unique among the spinel family [27]. The first NPD studies on $Co_2Ru_{1-x}Mn_xO_4$ provide evidence of change in magnetic order with Ru substitution and report two transitions at 100 and 180 K [25]. For lower Ru contents, a coexistence of long-range and short-range magnetic order was found where the complete Mn dilution results in the spin-glass-like ordering with spin freezing temperature of 16 K [25,28]. Further, in Bi-doped MCO, the magnetic transition increases significantly to 200 K [29]. Moreover, in a very recent study by Pramanik *et al*., the ferrimagnetic ordering was confirmed at 184 and 164 K in off-stoichiometric $Mn_{1.15}Co_{1.85}O_4$ and $Mn_{1.17}Co_{1.60}Cu_{0.23}O_4$ [30]. In another work, Pramanik *et al*. have confirmed the ferrimagnetic ground state of $Ti_{0.6}Mn_{0.4}Co_2O_4$ and $Ti_{0.8}Mn_{0.2}Co_2O_4$ based on neutron diffraction analysis below 110.3 and 78.2 K, respectively [31]. Enhanced distortion in $Ti_{0.6}Mn_{0.4}Co_2O_4$ as compared to $Ti_{0.8}Mn_{0.2}Co_2O_4$ was expected to be associated with higher $Mn^{3+}$ content in the later one [31].

In the present work, a detailed investigation of crystal and magnetic structure of $Ge_{0.5}Mn_{0.5}Co_2O_4$ (GMCO) has been performed. The end members GCO and MCO have cubic and tetragonal symmetry at room temperature, respectively [6,19,24]. This is because the JT active $Mn^{3+}$ ion in MCO enhances the structural distortion, resulting in tetragonal symmetry in MCO. Therefore, it will be interesting to investigate the effect of Mn substitution on structural and magnetic properties in GMCO. The Mn substitution is anticipated to change the crystal symmetry and related geometrical frustration, which has a significant impact on exchange interactions. Therefore, it will not affect the structural properties but will also alter the magnetic ordering/response, which is the main motivation for this work. Several interesting effects, such as re-entrant spin-glass, exchange bias, and tunning of ferrimagnetic temperature, have been observed with A and B site substitution, whereas only a single report with 20% Mn substitution at Ge site was recently published [32] based on bulk magnetization measurements. This work comprises details about the magnetic ground state of GMCO using neutron and x-ray diffraction (XRD), magnetization, and heat capacity technique and compares it with GCO. The synthesized compounds GCO and GMCO adhere to the similar normal spinel structure, characterized by a cubic space group *Fd-3m* at room temperature. However, at low temperatures, GMCO retains cubic symmetry, unlike GCO and MCO. Magnetic susceptibility indicates a ferrimagnetic transition at $T_N$ = 108 K. An additional antiferromagnetic (AFM) transition has been observed at further low temperatures, which is attributed to the minute impurity in the sample, which remains hidden in the XRD data. Heat capacity data has been analysed to confirm the nature of magnetic transition and to estimate associated magnetic entropy. Low-temperature NPD and XRD analysis has been carried out to investigate the effect of Mn substitution on crystal and magnetic structure, and the results are compared with low-temperature structural response of GCO. For magnetic structure determination, NPD data has been analysed in detail, offering insights into the spin structure of GMCO. Temperature dependence of average A and B site magnetic moments are determined based on Rietveld refinement. This work is focused on probing the detailed crystal and magnetic structure in $Ge_{0.5}Mn_{0.5}Co_2O_4$ and comparing it with other members of this family.

**Experimental**

The Polycrystalline $Ge_xMn_{1-x}Co_2O_4$ (x= 0 and 0.5) were prepared by a solid-state reaction method [4,15,18,21]. The starting materials, $GeO_2$ (99.99% purity), $MnO_2$ (99.99% purity), and $Co_2O_3$ (99.99% purity) were mixed in the stoichiometric amounts. The resulting mixture was then calcined at 950 °C for 12 hr and sintered at 950 °C for another 12 hr. To verify the phase purity of the prepared compounds and to investigate the low-temperature structural phase transitions, powder X-ray diffraction (XRD) measurement was recorded as a function of temperature using Rigaku's diffractometer equipped with Cu-K$\alpha$ radiation. The resulting diffraction data was Rietveld refined using JANA [33]. The DC magnetic susceptibility measurements were measured using a superconducting quantum interference device (SQUID) as a function of temperature and field, and the heat capacity was measured using a physical property measurement system. Powder neutron diffraction measurements on $Ge_{0.5}Mn_{0.5}Co_2O_4$ were carried out at ECHIDNA and WOMABT beamlines at the OPAL facility, ANSTO, Australia. The room temperature data from ECHIDNA beamline with neutron wavelength 1.62 Å was used to estimate the chemical composition of Mn substituted compound. Temperature-dependent data from WOMBAT beamline with neutron wavelength 2.41 Å was used for magnetic structure analysis as a function of temperature for a broad range of temperatures ranging from 5 to 120 K. Fullprof_suit was used for magnetic structure refinement [34,35].

**Result and Discussion**

Figure 1(a, b) shows the Rietveld refined XRD patterns $GeCo_2O_4$ (GCO) and $Ge_{0.5}Mn_{0.5}Co_2O_4$ (GMCO) recorded at room temperature. All the observed peaks are very well fitted with the cubic space group *Fd-3m*, consistent with the reported crystal structure of GCO and $Ge_{0.8}Mn_{0.2}Co_2O_4$ [32]. The XRD was refined by assuming the random distribution of $Ge^{3+}$ and $Mn^{3+}$ at the tetrahedral site 8*b*, where the $Co^{2+}$ is at octahedral site 16*c*. The oxygen is at 32*e*. The refined lattice parameters for GCO and GMCO are a = 8.3095(6) and 8.3085(6) Å. The inset in each figure enlarges the (008) reflection, which will split during the cubic to tetragonal transition at a lower temperature in GCO. No extra peak at the lower 2θ value was observed, which confirms that the Ge and Mn are randomly distributed at the A site. However, it is impossible to extract the exact cation distribution using XRD, and therefore NPD data have been refined for this purpose at room temperature.

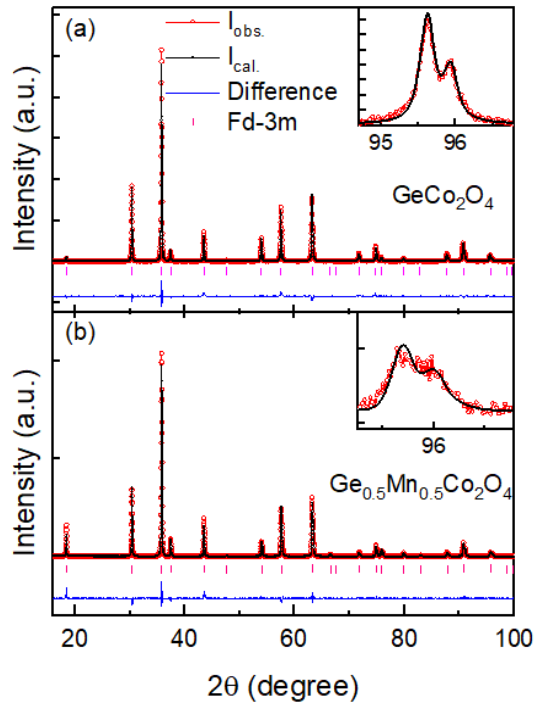

Figure 1: Rietveld refine XRD patterns of (a) $GeCo_2O_4$ and (b) $Ge_{0.5}Mn_{0.5}Co_2O_4$ using cubic space group $Fd$-$3m$.

Due to the substantial difference in the neutron scattering lengths for Mn (-3.73 fm) and Co (2.49 fm) ions, the room temperature NPD pattern from the ECHIDNA beamline was analyzed to determine the cationic distribution at the A and B sites in GMCO. The initial model, which assumed both $Ge^{4+}$ (50%) and $Mn^{4+}$ (50%) at the A site, failed to capture the full intensity of several reflections. Subsequently, refinement was iteratively improved through multiple steps, and the best fit with this model is illustrated in Figure 2(a). However, even after several iterations and checking all the possibilities of cation distribution at the A and B sites, this model could not capture the complete intensity of some of the reflections, as shown in the enlarged view provided in the inset of Figure 2(a). For the subsequent sections, this model will be denoted as model 1. To account for the full intensities of these reflections, an additional phase was introduced. The two-phase (model 2) refinement is given in Figure 2(b). After several iterative refinements, a reasonable fit was achieved, and the resulting cationic distribution in the main phase is estimated as $(Ge_{0.503}Mn_{0.44}Co_{0.06})(Co_{1.81}Mn_{0.19})O_2$, indicating the presence of Mn and Co at both A and B sites. The Oxygen and Ge occupancies were initially kept fixed and the relative occupancies of Mn and Co both at A and B sites were refined in several iterations. The value of $\chi^2$ reduced from 7.59 to 3.75, while the values of $R_{wp}$ reduced from 14.6 to 9.86 from model 1 to model 2. This indicates a significant improvement in the refinement, which is further evident from the inset of Figure 2(b). The refined lattice parameters for the primary phase measure a = 8.29642(7) Å, while for the secondary phase, they are slightly different at a = 8.30751(8) Å. Determining the precise composition of the secondary phase solely through refinement is not possible in this case. For refinement purposes, the second phase is simply assumed to be GCO, as the magnetic transition of the secondary phase matches with GCO, as explained in detail in the next section. The refined structural parameters of the first phase are given in Table 1.

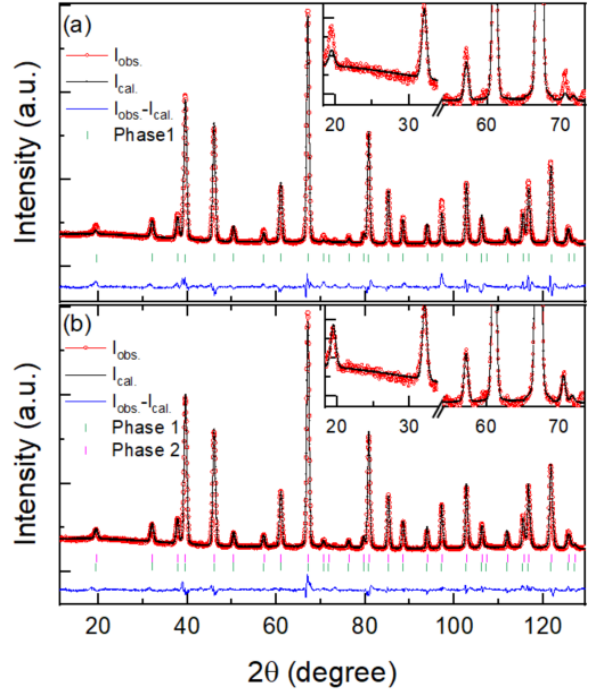

**Figure 2**: Rietveld refined room temperature powder neutron diffraction pattern of $Ge_{0.5}Mn_{0.5}Co_2O_4$ (a) single phase and (b) two-phase models. The insets in both figures depict the enlarged view around the 2θ value from 20 to 70 degrees.

Table 1: Refined structural parameters of GMCO from NPD data.

| Atoms (site) | x | y | z | B (Å²) |
|---|---|---|---|---|
| $Ge^{3+}/Mn^{4+}/Co^{3+}$ (8*b*) | 0.125 | 0.125 | 0.125 | 1.13(2) |
| $Co^{2+}/Mn^{2+}$ (16*c*) | 0.5 | 0.5 | 0.5 | 0.60(9) |
| O (32*e*) | 0.2533(1) | 0.2533(1) | 0.2533(1) | 1.22(1) |

Further to investigate the effect of Mn-substitution on the magnetic properties, DC susceptibility of GMCO has been measured and compared with GCO. Figure 3(a) shows the DC susceptibility-vs-temperature (χ–T) curves in the temperature range of 2 and 300 K measured with an applied field of 500 Oe after the zero-field cooled (ZFC) and field cooled (FC) conditions. The susceptibility sharply increases below 120 K and exhibits a peak around 95 K, accompanied by a bifurcation below it. From the susceptibility data, the curve indicates ferri/ferro ordering at this temperature. The transition temperature is estimated from the derivative which is $T_N$ =108 K. With further decreasing temperature, a small kink in the data can be seen around 22 K, which is more evident in the heat capacity data as shown in Figure 3(b). The susceptibility of pristine GCO is also shown in Figure 3(a), revealing a single transition at $T_N$ = 22 K, consistent with the literature [3,18,36]. Based on neutron analysis and comparing the data with GCO, it appears that this low-temperature transition in GMCO is originating from the secondary phase, GCO. Therefore, it is expected to be associated with antiferromagnetic (AFM) ordering of Co-spins in triangular and Kagome planes [12,13,26,37] which will be further discussed and confirmed using NPD analysis. The isothermal magnetization curves are the typical behaviour expected from a ferrimagnetic lattice with large anisotropy below $T_N$. Recently, Singha *et al*. have reported the magnetic properties of $Ge_{0.8}Mn_{0.2}Co_2O_4$ [32]. A longitudinal ferrimagnetic (FiM) order below 77 K due to uneven moments of divalent Co (↑ 5.33 $\mu B$) and tetravalent Mn (↓ 3.87 $\mu B$), coexists with transverse spin-glass state below 72.85 K was suggested [32]. However, it should be noted that in the absence of heat capacity and neutron data, it is very likely that a minute secondary phase that remains hidden in XRD data will be missed. The magnetic signal from this phase can be easily missed or misunderstood in the absence of complementary data as a second tranistion. The inverse magnetic susceptibility in the inset of Figure 3(a) exhibits deviation from CW behaviour below 200 K, which indicates that the magnetic interaction starts at a temperature far above $T_N$. Moreover, the value of Curie constant for GMCO is 6.04 which corresponds to the effective paramagnetic moment value of 6.95 $\mu_B$/f.u., consistent with the expected value. The expected $\mu_{eff.}$ value of GMCO lies between GCO and $MnCo_2O_4$ [21,24]. The value of $\theta_C$ is negative and equals -107.4 K for GMCO. The frustration index is 1.2, which indicates the absence of frustration in GMCO.

To further estimate the magnetic specific heat and magnetic entropy ($S_{mag.}$) during these magnetic transitions, the heat capacity of GMCO has been measured and presented in Figure 3(b). A broad transition with a peak centre at 108 K ($T_N$) is observed in the data. An additional sharp peak at 22 K is also evident. A sum of Debye and Einstein model is used to evaluate the lattice specific heat [38]. The estimated value of magnetic entropy changes equals 11.26 J/mole-K. Assuming $Mn^{4+}$ at A site and $Co^{2+}$ at B site, this value is nearly 63% of the expected value for GMCO. An alternative method based on harmonic lattice approximation was used by Lashley *et al*. [4] to estimate the magnetic entropy for GCO and $GeNi_2O_4$, which yield almost identical results. The reported values of magnetic entropy were only 58.3% and 56.5% in the case of GCO and $GeNi_2O_4$ and the missing entropy was suspected to be originating from substantial magnetic correlations well above $T_N$ [4]. A slightly higher value of entropy in our case could be due to the presence of additional transition at lower temperatures associated with the secondary phase. Moreover, the broad nature of high-temperature transition indicates the coexistence of short-range correlations in this compound, consistent with the recent results on $Ge_{0.8}Mn_{0.2}Co_2O_4$ by Singha *et al*. [32].

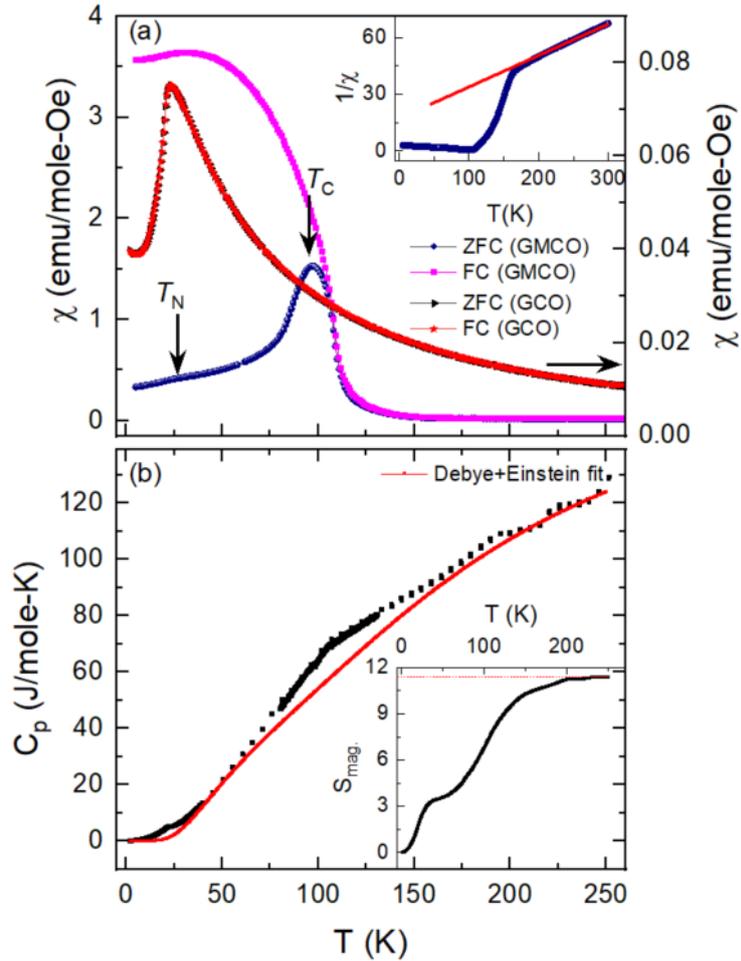

**Figure 3:** Temperature-dependent (a) DC susceptibility of GCO and GMCO and (b) the heat capacity behaviour of GMCO. The inset in (a) shows the CW-fit (red line) of the inverse susceptibility data for GMCO. The red line in (b) shows the fit using a sum of Debye and Einstein model. The inset in (b) shows the magnetic entropy change during both transitions.

Further, to investigate the effect of Mn substitution on low temperature crystal structure, XRD patterns in the selected 2θ range were recorded while cooling the sample from 25 to 12 K for both GCO and GMCO. Figure 4(a, b) shows the temperature dependence stack of (0 0 8) reflection. Clear splitting of (0 0 8) reflection starting from $T_N$ confirms the cubic to tetragonal ($I4_1/amd$) transition in GCO, associated with magnetic order. However, no such peak splitting was observed for GMCO, confirming the cubic symmetry down to 12 K. Regarding the splitting in GCO, it is clearly evident that the splitting of (0 0 8) peak can be seen starting from $T_N$ (= 21 K). Earlier, Barton *et al*. [19] argued that the structural transition in GCO is decoupled from the magnetic transition based on powder data. However, in the present case, it occurs around the same temperature, indicating a strong coupling between structural and magnetic order. Similar distortion at the Néel temperature in CoO was under debate, while some reports indicate spin-orbit couple magnetostriction [3,39] due to degenerate $t_{2g}$ states in octahedral $Co^{2+}$, and others suggest Jahn-Teller ordering [40,41]. Moreover, these distortions can be suppressed or decoupled from the magnetic ordering, resulting in the onset of AFM ordering without any accompanying structural transition [40]. In the case of GMCO, no splitting or broadening of (0 0 8) was observed. This indicates that with the added disorder at A site, the degeneracy is lifted, resulting in the magnetic order without any accompanying distortion. The origin of structural distortion in these systems with degenerate $t_{2g}$ states could have multiple origins or coupled effects of spin, orbital, and lattice degrees of freedom. Due to the complex cation distribution in GMCO and several possible factors contributing to distortions in these compounds, it is challenging to pinpoint the specific reason for the observed effect.

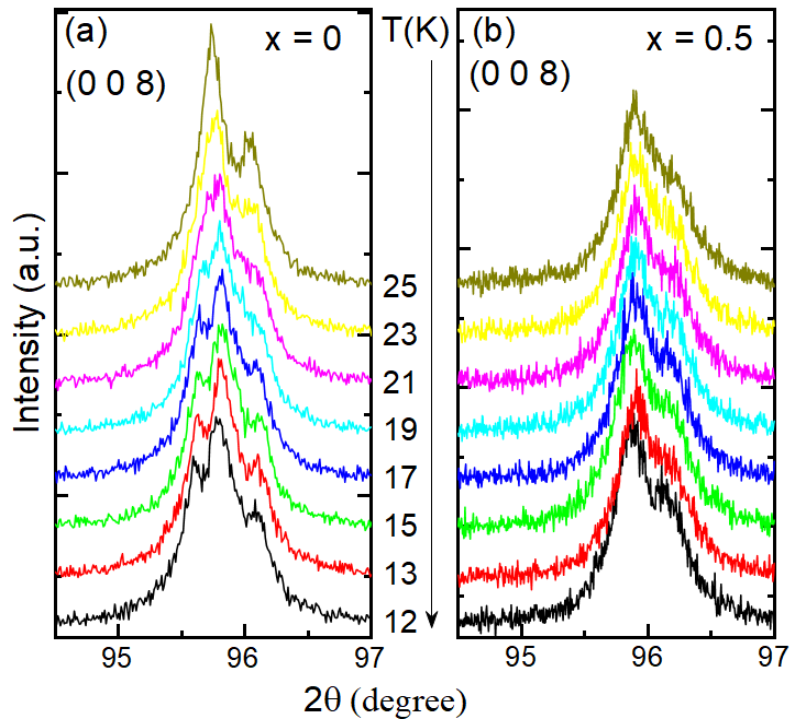

**Figure 4**: Stack of XRD profiles of (0 0 8) reflection for (a) GCO and (b) GMCO.

Furthermore, full XRD patterns have been recorded at 12 K for both compositions. Figure 5(a-b) shows the Rietveld refined XRD pattern of GCO and GMCO, collected at 12 K. The insets in each figure enlarged the view of (0 0 8) peak. The refined lattice parameters at 12 K for GCO are 5.8777(1) and c = 8.3004(1) whereas for GMCO, the cubic lattice parameter is a = 8.29658(9) Å. Sometime in case of weak distortions, peak broadening has been observed for other members of this family [30]. However, in the present case no significant broadening with temperature was observed for GMCO. The full width at half maximum (FWHM) of the (0 0 8) reflection decreases from room temperature to 12 K by the value of 0.322(1) to 0.319(9)°, confirming the absence of structural phase transition in GMCO.

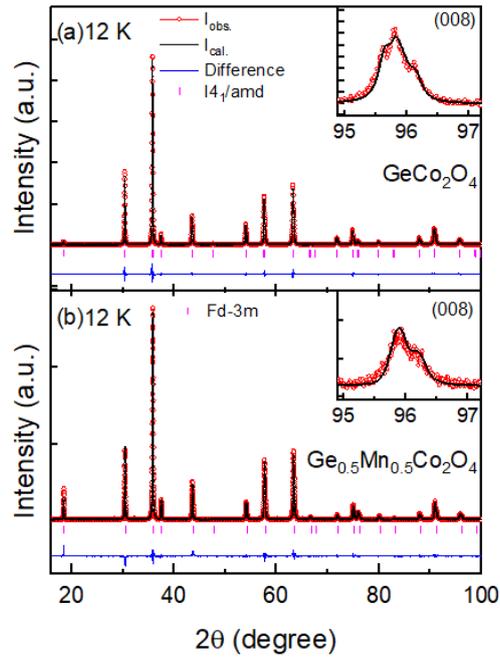

**Figure 5**: Rietveld analysis of XRD pattern of (a) $GeCo_2O_4$ and (b) $Ge_{0.5}Mn_{0.5}Co_2O_4$ measured at 12 K.

For magnetic structure determination, NPD patterns have been collected on the WOMBAT instrument for a temperature range of 160 to 5 K. Figure 6(a) presents the NPD patterns at the selected temperature, measured with $\lambda$ = 2.41 Å. The difference curves for 30 and 5 K with respect to 160 K are depicted in Figure 6(b). While cooling the sample below $T_N$, the intensity of some of the nuclear reflections increases significantly at lower $2\theta$ values, indicating ferro/ferrimagnetic ordering. The integrated intensity of (1 1 1) reflection highlighted within the box region in Figure 6(a) is plotted in the inset of Figure 6(b). An increment in the intensity of (1 1 1) can be seen below 130 K, which is slightly above $T_N$. These magnetic reflections for temperature range 25 K<T<$T_N$ can be indexed with a propagation vector (0, 0, 0). As the temperature further decreases, several new peaks emerge adjacent to nuclear reflections below 22 K, originating from the second phase. The difference data of 160 to 5 K in Figure 6(b) reveals these additional reflections more clearly below 25 K. These additional magnetic reflections can be indexed with propagation vector (1/2, 1/2, 1/2), consistent with the AFM ordering of Co ions in GCO [21,26,30].

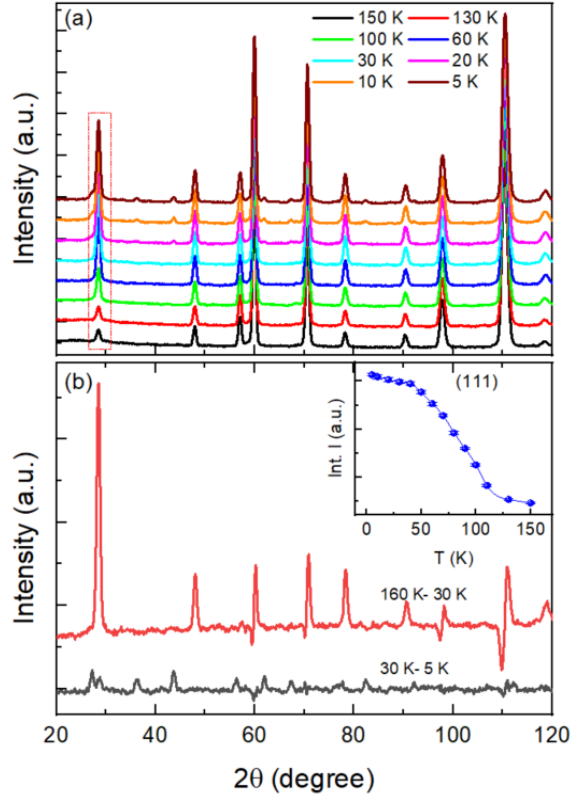

**Figure 6:** NPD patterns for $Ge_{0.5}Mn_{0.5}Co_2O_4$ recorded at (a) selected temperatures and (d) difference patterns reveal the two sets of magnetic diffraction. Incet to Fig. 6(b) shows the temperature dependence of (1 1 1) peak.

To determine the magnetic structure at different temperatures, refinements have been performed with single nuclear (model 1) as well as two nuclear phase (model 2) models. Figure 7 compares the magnetic refinement at 30 K using two models, model 1 and model 2. In model 1, a single nuclear phase and single magnetic phase were used whereas in model 2, two nuclear phases and one magnetic phase were used. Although the quality of refinement improved with model 2, it is hard to judge by comparing the refined patterns with eyes. The $R_{wp}$ values for models 1 and 2 are 9.7 and 7.48, respectively, whereas the $\chi^2$ values are 9.82 and 6.15, respectively. Further, in both these models, the A and B-site moment is refined by assuming a random distribution of Co and Mn at A and B sites with equal moment values. Also, the occupancies and scale factor were kept fixed from 160 K refinement. The refinement didn't converge when the Co and Mn moment at A or B site were not constrained to be the same. The resultant magnetic structure is shown in Figure 8(a). The resultant magnetic structure has ferrimagnetic coupled A and B site moments which are aligned parallel to *c*-axis. The average A and B site moments are plotted in Figure 8(b). The values of average A and B -site moments are consistent with those reported for $Ti_{1-x}Mn_xCo_2O_4$, $Mn_{1.17}Co_{1.60}Cu_{0.23}O_4$ by Pramanik *et al*. [30,31]. The estimated phase fraction of the second nuclear phase is 7.7 %. The presence of this second minority phase doesn't affect the overall magnetic structure. However, the moment value changes slightly with model 2. Table 2 compares the value of A and B site moments using two models at 30 K.

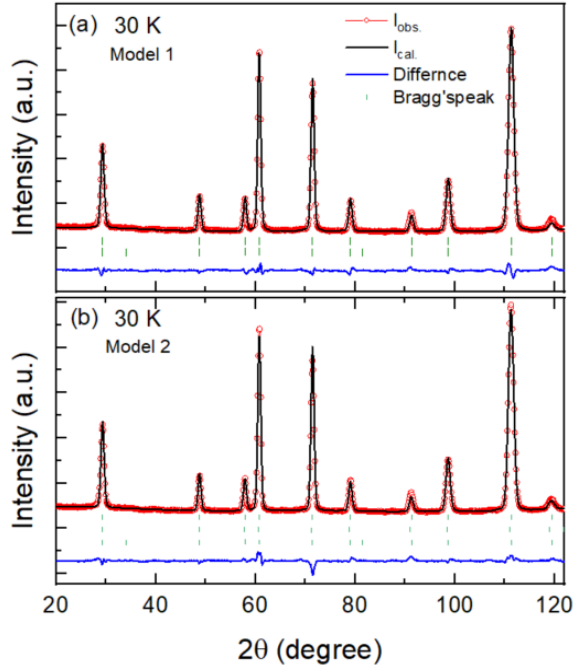

**Figure 7**: Rietveld refined NPD pattern at 30 K using (a) model 1 (single phase) and (b) model 2 (two phases). For more details about different models, please check the text. The Bragg peaks from top to bottom in (a) represent the GMCO nuclear model, the magnetic model from GMCO, and in (b), they represent the GMCO nuclear model, GCO nuclear model, and GMCO magnetic model, respectively.

Table 2: Magnetic moments of the transition metal atoms in $Ge_{0.5}Mn_{0.5}Co_2O_4$ obtained from Rietveld refinements using neutron diffraction data collected on WOMABT beamline at 30 K using two models.

| Magnetic moments ($\mu_B$) | Model 1 | Model 2 |
| --- | --- | --- |
| $\mu(Mn_A)$ | 2.53(4) | 2.29(4) |
| $\mu(Co_B/ Mn_B)$ | 1.51(3) | 1.79(3) |

Moreover, magnetic refinement at different temperatures was performed to estimate the average A and B site moments as a function of temperature using model 2. Figure 8(b) shows the refined moment values as a function of temperature using model 2. From Figure 8(b), it is evident that both the A and B site moments increase and attain the maximum value of around 20 K, which remains almost constant further down to 5 K.

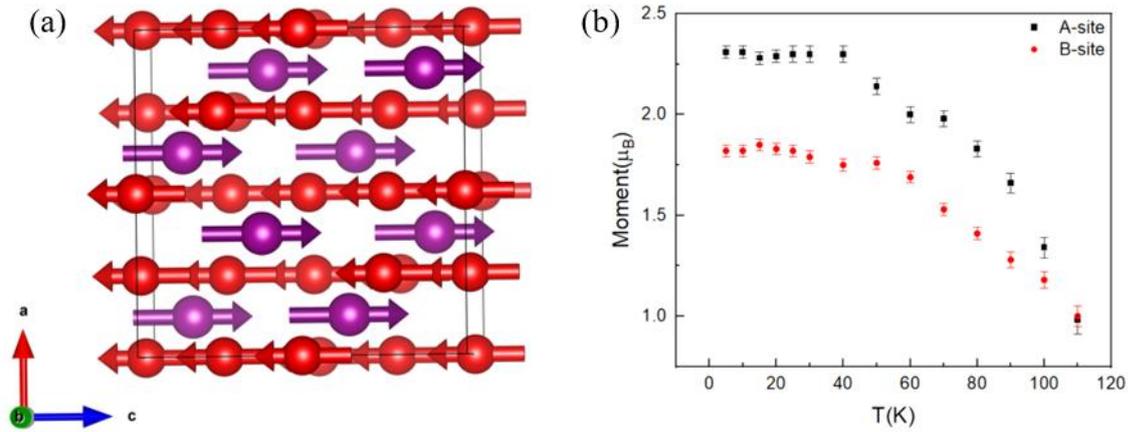

**Figure 8**: (a) Magnetic structure at 30 K. (b) Average A and B site magnetic moment obtained from refinement. The red and violet atoms denote A and B-sites, respectively.

For the magnetic structure refinement below 25 K in the secondary phase, an additional magnetic phase with propagation vector (½, ½, ½) was added in the pcr file to account for the extra magnetic peaks appearing below the second transition. The resultant magnetic structure in this second magnetic phase is consistent with existing reports on GCO, having alternate stacking of Kagome and triangular planes along the [1 1 1] direction. All the extra peaks observed below 22 K can be indexed with this model. The refined pattern at 5 K is shown in Figure 9. Also, for the refinement this secondary phase, the scale factor is kept same as that of second nuclear phase. The average B site moment at 5 K in this secondary phase is 2.53(5) $\mu_B$ which is consistent with literature [31]. The inset in Figure 9 exhibits the temperature dependence of $Co^{2+}$ moment in this secondary phase. Overall, a secondary phase with estimated 7.7 % phase fraction is associated with this AFM ordering below 22 K.

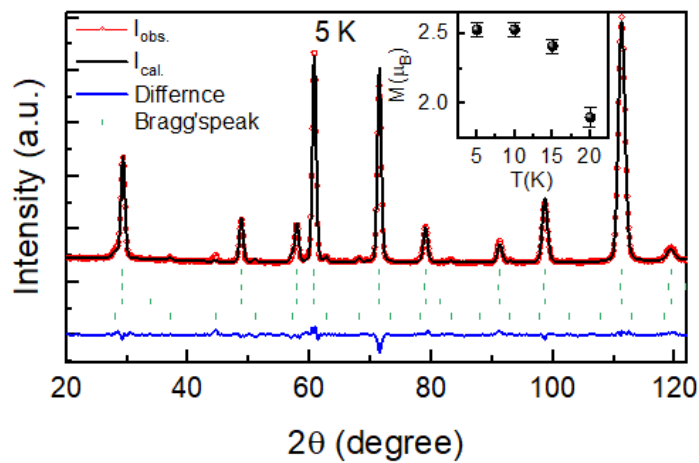

**Figure 9**: Rietveld refined NPD pattern at 5 K using two magnetic phases. The inset in (a) shows the $Co^{2+}$ moments as a function of temperature. The Bragg peaks from top to bottom in the plot denote the GMCO nuclear phase, GCO nuclear phase, GMCO magnetic phase, and GCO magnetic phase, respectively.

**Conclusion**

A comprehensive study of the structural and magnetic behaviours of $Ge_{0.5}Mn_{0.5}Co_2O_4$ (GMCO) has been conducted, revealing the effects of Mn substitution in these complex oxide materials. The findings were compared with those of $GeCo_2O_4$ (GCO) to provide a deeper understanding of the influence of added disorder on crystal and magnetic structure. Based on the detailed analysis of XRD and NPD data, it is confirmed that both GCO and GMCO crystallize in a cubic structure with slight difference in lattice parameters at room temperature. Interestingly, even though the room temperature structure is nearly identical of both compounds, the low temperature structural and magnetic response has significant differences. Bulk magnetization and heat capacity data indicates two magnetic transitions in GMCO at 108 and 22 K, unlike GCO with single AFM transition at 21 K. Moreover, the magnetic entropy changes in GMCO suggest significant magnetic correlations above the transition temperature. Further the low temperature XRD results confirms a cubic-to-tetragonal transition in GCO associated with magnetic order, which is absent in GMCO, indicating that disorder at the A and B- site suppresses structural distortions linked with magnetic transitions. In general, it is non-trivial to pinpoint the specific effect responsible for coupled magneto-structural transition but in the present case, it appears that with Mn substitution, the structural distortions with degenerate $t_{2g}$ states are suppressed or decoupled from the magnetic ordering in GMCO. Using Rietveld refined NPD data, the cation distribution in GMCO has been determined which indicates the presence of both Co/Mn at A and B sites. Further the NPD analysis reveals a secondary phase resembling GCO in GMCO, likely due to incomplete substitution or segregation at the A site. The low temperature transition at 22 K in GMCO is confirmed to be associated with minor secondary phase which remains hidden in the XRD data. Magnetic structure analysis of GMCO confirms the ferrimagnetic order below 108 K associated with the main phase. The A and B-site moments are coupled ferrimagnetically along the $c$-axis. Magnetic structure analysis of secondary phase is consistent with the magnetic structure of GCO reported earlier by several groups. This work provides critical insights into how Mn substitution or the added disorder influences the magneto-structural properties of spinels, enhancing our understanding of complex oxide materials.

**Acknowledgements**

A portion of this work was performed at the National High Magnetic Field Laboratory, which is supported by National Science Foundation Cooperative Agreement No. DMR-2128556 and the State of Florida. B.S. and K.W. acknowledge the support of the NHMFL User Collaboration Grant Program.